# Impersonation Attack using Quantum Shor's Algorithm against Blockchain-based Vehicular Ad-hoc Network

Kazi Hassan Shakib, *Student Member, IEEE*, Mizanur Rahman, *Senior Member, IEEE*, Mhafuzul Islam, and Mashrur Chowdhury, *Senior Member, IEEE*

*Abstract*—Blockchain-based Vehicular Ad-hoc Network (VANET) is widely considered a secure communication architecture for a connected transportation system. With the advent of quantum computing, there are concerns regarding the vulnerability of this architecture against attack algorithms implemented in a quantum computer. In this study, a potential threat is investigated in a blockchain-based VANET with an impersonation attack utilizing Shor's algorithm implemented in a quantum computer. Specifically, the impersonation attack using Shor's algorithm is created by compromising the Rivest-Shamir-Adleman (RSA) encrypted digital signatures of VANET, and thus a threat to the trust-based blockchain scheme of VANET is successfully established. We implemented an integrated simulation platform combining OMNET++, vehicles-in-network simulation (VEINS), and simulation of urban mobility (SUMO) traffic simulator. In addition, we incorporated vehicle-to-everything (V2X) communication in OMNET++ using the extended INET library. An impersonation attack on a blockchain-based VANET is implemented using IBM Qiskit, which is an open-source quantum software development kit. The findings reveal that an impersonation attack is feasible on the Blockchain-based VANET, which compromises the trust chain of a blockchain-based VANET. This research highlights the need for a quantum-secured blockchain for VANET.

## I . INTRODUCTION

### A. Background

RECENTLY, blockchain-based Vehicular Ad-hoc Network (VANET) architectures have gained significant attention due to its distributed and decentralized architecture [1], providing efficient data transmission capability with secure data generation and broadcasting ability over VANET networks [2]. These architectures often employ rating-based or trust-value-based mechanisms, utilizing consensus mechanisms, such as proof-of-work (PoW) and proof-of-stake (PoS), to ensure trustworthiness across a communication network [2]. Such a trust management system could ensure secure and privacy-protected [1] vehicle-to-everything (V2X) communication because of its ability to maintain the veracity of the exchanged messages via a digital signature of a message sender (e.g., vehicle). RSA (Rivest-Shamir-Adleman), a widely used public-private asymmetric key cryptography technique, is commonly utilized in blockchain-based VANET architectures to encrypt messages.

Given the low-latency data transmission requirements in VANET, minimizing delay in transmitting data packets is necessary for real-time vehicle-to-vehicle (V2V) communication. So, to satisfy communication latency requirements, a small key-based encryption is necessary, which is recommended and commonly used in VANET [3], [4]. Prior studies support the use of small key based encryption for blockchain security as it requires less complex computational operations and storage [5], [6].

Given the limited computing resources of current quantum computers, this study demonstrates an impersonation attack on VANETs by scaling down the cryptographic standards (i.e., key size) of Blockchain. The findings highlight vulnerabilities under existing quantum resource constraints, proving the feasibility of an impersonation attack on Blockchain-based VANETs utilizing Shor's algorithm implemented in a quantum computer. Furthermore, the results indicate that with sufficient quantum computing resources, the same attack scheme could compromise scaled-up cryptographic standards in Blockchain-based VANETs.

### B. Potential Attack and Its Impact on Blockchain and VANET

A classical attack model, such as a false message by any attacker, will cause a lack of trust in V2V communication, and a blockchain-based VANET can identify such malicious vehicles using a trust-based blockchain framework. However, there are cases where the trust-based framework of blockchain could become vulnerable. For example, a blockchain-based VANET cannot identify a malicious vehicle and corresponding false messages in an impersonation attack, as shown in Figure 1. In this scenario, a group of vehicles, where each vehicle acts as an agent node [1], creates a blockchain under a Roadside Unit (RSU), and the RSU runs a trust management unit. All RSUs between each other form (as shown between #(N+1) to #(N+2) RSU) a trust evaluation chain, which is a chained framework to store and transmit trust information. As presented in Figure 1, a malicious vehicle (orange in color) could impersonate a legitimate vehicle (green in color) by forging its digital signature using a quantum Shor's algorithm. The attacker could generate an impersonation message (Message #1) of a false traffic crash and broadcast the false message to surrounding vehicles, as shown in Figure 1. All vehicles within its communication range, upon receiving the message (Message #1), send the acknowledgment (ACK) message (Message #2) to RSU #(N+1). RSU #(N+1) then disseminates the transaction message (Message #3) to other vehicles for their consensus regarding the transaction. After that, if it is an authenticated and trusted message through the consensus messages (Message #4) of the vehicles of this local network, RSU #(N+1) sends this authenticated false crash message (Message #5) to all other vehicles and RSUs within its communication range. In this scenario, RSU #(N+1) cannot find the malicious (original) sender (orange in color) of the false traffic crash message, as the digital signature of the legitimate (green) vehicle has been compromised by the malicious vehicle (orange in color). So, the trust mechanism fails to detect the attacker's vehicle. Furthermore, if the attacker's vehicle receives the miner/validator capability—i.e., writing capability on the new blocks—as a legitimate vehicle, it also validates another malicious vehicle's false message in the blockchain and can create a favorable scenario for the attacker. This scenario shows an example of the vulnerability of a blockchain-based VANET against an impersonation attack.

The impact of this type of attack on a blockchain-based Vehicular Ad-hoc Network (VANET) extends beyond merely disrupting; it can lead to severe consequences, such as roadway traffic crashes and significant congestion. Thus, false information about traffic incidents may misguide vehicles and traffic management systems and cause accidents or delays. Moreover, the compromise of the trust mechanism within the VANET's blockchain allows malicious actors to influence the operations of vehicles in a transportation network. This influence can lead to the validation of false transactions, including the introduction of fraudulent activities. The infiltration of false transactions undermines the reliability and integrity of the blockchain, eroding trust in the communication network's ability to accurately record and validate transactions. Consequently, the entire VANET ecosystem becomes vulnerable to manipulation and disruption, putting at risk its capacity to facilitate secure vehicular communication and coordination.

### C. Assumptions

The vulnerability in the trust-based framework of blockchain arises in scenarios like impersonation attacks, where a malicious vehicle forges digital signatures of legitimate vehicles and compromises the trust mechanism in a vehicular network. Given the low latency communication requirements within a dynamic VANET environment, small key-based encryption is required. However, in this paper, the decision to use the small key size of cryptographic schemes within the blockchain technology is because of the current constraints of quantum computing. The study investigates the feasibility of an impersonation attack on a small key-based blockchain-based VANET and presents a proof-of-concept demonstrating the potential of an impersonation attack utilizing Shor's algorithm implemented in a quantum computer. The experimental scenario is limited to four vehicles and one RSU to serve as proof of concept. The vehicles provide consensus, while the RSU acts solely as a local administrator without participating in the voting process. Thus, a constraint on vehicular network size does not impede the validity of the proof of concept, as it demonstrates the interoperability of the trust chain within VANET. This impersonation attack is not feasible with classical computing but becomes feasible with quantum computing. Therefore, if the encryption key size is increased, the same impersonation attack scheme would be feasible, provided there is sufficient quantum computing capability. These assumptions guide the objectives of the paper, which focus on analyzing blockchain vulnerabilities, developing a proof of concept for an impersonation attack utilizing Shor's algorithm implemented in a quantum computer, and validating the impersonation attack model to assess potential threats in VANET architecture with the increasing power of quantum computers.

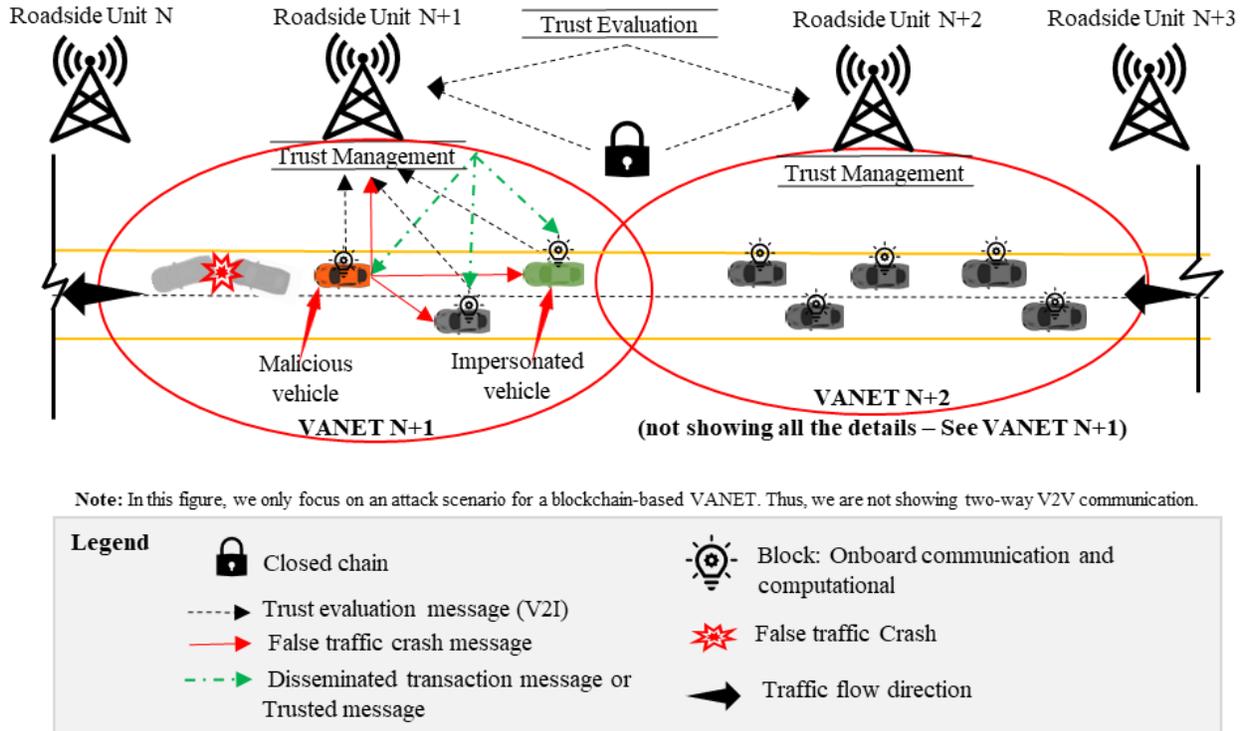

**Fig. 1.** An example attack scenario for a Blockchain-based VANET.

*D. Objectives and Contributions*

A blockchain-based architecture relies on two cryptographic mechanisms to provide security and trust [7]: (i) check the integrity of the data itself using hash functions, and (ii) check the ownership of the data with asymmetric cryptography. If an attack algorithm implemented in a quantum computer can compromise the cryptographic algorithm, it can create security concerns for any secure communication architectures, such as blockchain, as it uses an encryption technique (mostly on subgroup-finding algorithms utilizing factorization and discrete logarithm), i.e., RSA [8], elliptic curve digital signature algorithm (ECDSA) [3], [9]. In this paper, we developed an impersonation attack framework utilizing Shor's algorithm, which was implemented in a quantum computer, and exploited the vulnerability of the cryptographic ownership mechanism. The objectives of this study are to:

(a) analyze vulnerabilities of secured blockchain-based VANET,
(b) develop and validate an impersonation attack model utilizing Shor's algorithm implemented in a quantum computer to break the trust-based architecture of a blockchain-based VANET that uses asymmetric encryption, such as RSA and ECDSA.

This proof-of-concept demonstrates a potential threat of attacks using quantum computers on the public-private key-based cryptography in VANET architecture. Findings from this study indicate existing vulnerabilities of blockchain-based VANET architecture and provide insights into potential threats that could arise due to quantum computing advancements.

## II. LITERATURE REVIEW

For non-quantum-based attacks, prime factorization, which is a major part of cryptography used in ownership mechanism, cannot be broken in polynomial time [10]. Note that modular exponentiation is one of the necessary steps in prime factorization. However, modular exponentiation cannot be executed in polynomial time by classical computing algorithms, and it requires sub-exponential time[11]. Because of the superposition property of qubits, quantum computers can execute modular exponentiation in sub-exponential time. For example, it is possible to do factorization in polynomial time with a quantum computer, where a classical computer needs sub-exponential time to perform prime factorization [10]. Due to the nature of finding non-trivial factors in every iteration, Shor's algorithm estimates and is more likely to find factors in a polynomial time [10].

Quantum attacks can be formulated using two different algorithms implemented in a quantum computer: (i) Shor's algorithm [10] because of its ability of factorization; and (ii) Grover's algorithm due to its searching capability [12]. The Shor's algorithm provides a threat on asymmetric cryptography through prime factorization and discrete logarithm problem-solving capability, which covers the basis for a wide range of cryptographic techniques. On the other hand, Grover's algorithm is applicable for symmetric cryptography through inversing the hash functions [13]. In this study, we have leveraged the efficacy of Shor's algorithm for developing an impersonation attack scheme due to our focus on PoS based architecture and our impersonation attack. The Practical Byzantine Fault Tolerance (PBFT) [14], [15] one of the popular consensus algorithms for blockchain-based VANETs, could be vulnerable to quantum attacks. The PBFT relies on digital signatures and secure message exchanges to maintain trust and consensus in private or semi-decentralized networks. However, Shor's algorithm could pose a similar threat by compromising the authenticity of messages as presented in our paper.

Grover's algorithm is more appropriate for PoW-based scheme. The authors in [7] presented two primary applications of Grover's algorithm. Firstly, the algorithm can enable quadratic speedup for solving hash functions and creating attacks [16]. The attacker, who controls more than half of network's mining rate as miner, can monopolize the data and even rewrite the blocks by creating forks as well. Secondly, the attacker can search for hash collisions [16], [17] to replace certain parts of blocks without breaking the chain. Moreover, the attacker can construct a more reliable chain by maintaining the hash pointer of the block identically, as the pointer of the previous block will then reference the same block. Currently, researchers explore the complexity of securing communications in VANETs, emphasizing the challenges posed by the highly dynamic and fast-paced nature of vehicle interactions. Traditional approaches, such as public key infrastructure (PKI) [18], and identity-based signature schemes[19], face issues related to certificate management, key escrow problems, and computational overhead. Various attempts, including certificateless public key cryptography (CL-PKC) [20], [21]and certificateless aggregate signature schemes (CLAS)[22], certificateless lattice-based schemes like CLLS[23], aim to address these challenges but encounter heavy computation power and communication latency. To enhance efficiency in CL-PKC, Mei et al. [22]presented a CLAS scheme, which relies on bilinear pairing and map-to-point hash functions, demanding resources beyond the capacity of onboard units (OBUs). In another study, Ali et al.[19] incorporated blockchain to enhance revocation transparency, allowing participants to efficiently verify revoked pseudo-identities without relying on a trusted authority (TA). Despite this, their scheme's efficiency is compromised due to the necessity of a bilinear pairing during signature verification.

On the other hand, attacks that do not utilize quantum computers, such as Sybil attacks[24], one can exploit weaknesses in cryptographic mechanisms that use bilinear pairings, such as identity-based encryption schemes [25]. Since bilinear pairings allow operations across different groups, an attacker can manipulate inputs to link multiple fake identities to a single key, tricking the network into recognizing multiple Sybil nodes as legitimate entities. To counter such threats, recent studies have proposed blockchain-driven authentication protocols that ensure secure data exchange between vehicles and infrastructure, mitigating risks associated with malicious activities. For example, handover authentication mechanisms in these protocols[26], [27], [28] use smart contracts to automate authentication during handover, ensuring trust without significant latency. Blockchain systems also facilitate key exchanges by assigning temporary identities to vehicles and periodically changing cryptographic keys, which enhances security against replay and impersonation attacks. However, all these studies rely on the complexity of cryptographic solutions and their inherent property of computational infeasibility, which ensures that adversaries cannot feasibly reverse-engineer keys or signatures without incurring prohibitive computational costs, thereby maintaining secure communication even in the presence of malicious actors.

To ensure such secure communication, blockchains typically rely on robust digital signature methods for encryption, such as ECDSA, known for its strength but demanding considerable computational time. For instance, the Ethereum blockchain utilizes ECDSA, which takes nearly 10 milliseconds (ms) for a single signature and verification process [29], this would exponentially grow with aggregate signature schemes. To mitigate this delay, a lightweight encryption algorithm like RSA-1024 [30], [31] was employed, offering 80-bit security strength and requiring only one-third of the time compared to ECDSA-RSA-1024. The authors in[32] describe a transaction hijacking scenario in cryptocurrency platforms using the public key, which is published to the network or revealed from transactions sitting in the memory pools. They present an attack model that uses the Shor's algorithm on a known public key to decrypt the private key. Then, using that decrypted private key creates a conflicting transaction spending the same value. They created a commit-delay-reveal architecture to safeguard the reveal of public key encryption. However, such a scheme can fail in a broadcasting scenario, such a delay will cause a lack of trust. Another limitation of the study was the failure to provide proof-of-concept and the effectiveness of such attacks with experiments or simulations.

In another study [33], the authors explored vulnerabilities of five cryptocurrencies and their variations: Bitcoin, Ehereum, Litecoin, Monero and ZCash. Among these, Bitcoin exhibited vulnerability to duplicate transactions due to its signature scheme, while Ethereum faced potential replication issues with its consensus-based approach relying on digital logarithm-based public keys. Similarly, Litecoin shared vulnerabilities with Bitcoin due to its analogous structure. Monero employed ECDSA and a bulletproof protocol to hide transaction amounts, while ZCash, which uses consensus-based scheme, utilized ECDSA. The focus was on assessing the susceptibility of these cryptocurrencies to quantum attacks targeting their signature schemes and consensus algorithms. The study revealed vulnerabilities against transaction forging and duplication attacks. While offering theoretical proof of the cryptocurrencies' vulnerability against algorithms like Shor's and Grover's, the authors do not provide experimental analysis to quantify these vulnerabilities with a comprehensive attack model. In contrast, the presented method in this study not only identifies vulnerabilities within blockchain and cryptocurrency-based systems but also presents both comprehensive experimental analysis and an attack strategy against PoS-based blockchain schemes, along with a generic attack model applicable to PoW-based blockchains.

The authors in [34] discussed attacks using quantum computers on Bitcoin and potential countermeasures to protect against them. Their research specifically focuses on the impact of quantum attacks on Bitcoin and other cryptocurrencies. They examined the susceptibility of Bitcoin's PoW to quantum attacks, considering the public key disclosure through address reuse, along with the vulnerability of the cryptographic ECDSA algorithm employed in Bitcoin transactions. They found that while the current specialized application specific integrated circuit (ASIC) miners make the PoW relatively resistant to quantum speedup, the ECDSA signature scheme is more vulnerable and could be broken by a quantum computer as early as 2027. The study explores a detailed analysis of the time required for quantum computers to perform attacks on hash function using Grover's algorithm and compares the estimated quantum computing power against the total hashing power of the Bitcoin network. It also explores alternative PoW schemes and reviews various post-quantum signature schemes as countermeasures. However, as this is theoretical proof of concept only, it lacks comprehensive analysis.

Existing literature lacks a practical attack model that uses quantum computers to test (without only theoretical analysis) against the traditional asymmetric cryptographic schemes and lightweight key structure feasible for dynamic VANET scenarios. This indicates the importance of this research and the gap in existing literature.

## III. BLOCKCHAIN-BASED VANET AND THREAT MODEL

This section presents an architecture of blockchain-based VANET and identifies its vulnerabilities against an attack that uses quantum computers.

### A. Blockchain-based VANET Architecture

The blockchain-based VANET architecture has not been standardized yet. However, two different mechanisms, i.e.,

PoW and PoS based consensus [1], [2], are available in literature for implementing VANET architecture. Unlike PoW, which requires the computational power for solving cryptographic puzzles (i.e., solving SHA-X), PoS considers all the vehicles as stakeholders of the system, and trustworthy vehicles or randomly selected vehicles to be considered as the 'Miner' criteria. In PoW and according to the consensus mechanism, a vehicle sends a traffic crash warning message, and it goes to the "Mempool" of the RSU/Miner. The Mempool is where the unauthenticated messages or transactions stay as long as there are no similar reports from other vehicles regarding the message [35]. On the other hand, the consensus mechanism of PoS checks the traffic crash message with other vehicles' (stakeholders') consensus/opinion about the message and finally validated by validators/miners. PoS can find malicious transaction through horizontal trust (HT) in V2V communication and vertical trust (VT) in V2I communication [36].

The trust-chain is a trust-based blockchain architecture with asymmetric encryption to identify ownership using the digital signatures of the vehicles and their trust values, which increase with every authenticated transaction. A trust chain-based blockchain architecture verifies authenticity in two steps. Firstly, it verifies the ownership of the digital signature that is encrypted with RSA/ECDSA. Secondly, it checks the authenticity of the message using the consensus mechanism. As presented below in Table I, system configuration can be defined from the enrollment, transaction handling, consensus mechanism, storing the authentic request to blockchain record, miner/validator selection, trust value calculations and blocking of malicious vehicles after some transactions, which act as fault tolerance threshold. Based on the above definitions, attacks can be defended through authentication, transparency and non-repudiation nature of blockchain networks. For example, a vehicle with vehicle ego-motion information with false trajectories (e.g., speed-difference and distance), false digital signature, or false information about a traffic crash can be identified using a consensus mechanism.

### B. Threats and Vulnerabilities of a Blockchain-based VANET

The threats to a blockchain-based VANET system can be considered low if the trustworthiness of the system is maintained. Tempering of data and sniffing in ongoing VANET conversations (V2V and/or V2I) are considered less possible because of trustworthy encryption, such as RSA, hashing of the data, and distributed transparent ledger. However, the digital signature of a vehicle, which is part of each disseminated message, can be used to create an attack algorithm that uses quantum computers and makes a trust-based system vulnerable and can have a catastrophic impact on VANET, even creating the possibility of multi-vehicle collisions and changing the routes of vehicles. As illustrated in Figure 1, a general attack scenario on the PoS mechanism is presented, where a malicious vehicle generates a false message. Through the blockchain's consensus mechanism, the malicious activity is detected, tracked, and ultimately blocked. Then, a digital signature from disseminated packets of RSU is used to forge the digital signature of a legitimate vehicle.

TABLE I
COMPONENTS AND THEIR CORRESPONDING FUNCTIONALITIES FOR BLOCKCHAIN BASED VANET

| System architecture components | Description |
| --- | --- |
| Enrollment to the blockchain network | System interactions with blockchain network is initialized by all the agent nodes (vehicles as moving nodes and RSU's act as static nodes) downloading and updating following the architecture presented in [37] and the blockchain as a module (as Certificate Authority based authentication is not needed for PoS based architecture) with digital signature generated by RSU. Current blockchain transaction history has also been downloaded with respective trust values when a new vehicle joins the architecture |
| Transaction handling | There is a forward dissemination of packets from RSU (local administrator) to all vehicles within its wireless communication coverage to let every other vehicle know about any transactions and corresponding trust values to vote for the transaction authenticity. |
| Consensus mechanism | Voting is conducted by horizontal trust result (trusted=1 and not trusted=0) created by every transaction, which is sent to RSU (as VT step) for consensus. RSU calculates majority voting (greater than the threshold trust value) for every transaction and classifies malicious transactions. This threshold changes with increasing trust of the vehicles. |
| Blockchain record | Verified transactions with the approval of stakeholders are stored in the blockchain as a new block by the validator and dissipated to every agent node to maintain transparency in local blockchain module. |
| Miner/validator selection | To generate a block, entities called miners/validators (special agent nodes) with high trust value verifies in accordance with RSU's trust result, writes verified data to new block of blockchain in accordance to defined rule (threshold in voting or trust value calculation) through smart contracts. A vehicle with the highest trust value (the highest stake value) or a randomly selected vehicle (if many vehicles have the same trust value) serves as a validator/miner in PoS. |
| Trust value calculation | This trust value offset has been calculated as described in the reference [2]. Trust value can be increased with every vehicle authenticity check and message validation (adds 5 per transaction in this paper), with identifying false transactions (adds 10 per transaction), and with adding new validators/miners (20 for mining) upon writing onto the block. If a malicious vehicle tries to do double spending with the same gas value (i.e., compensation value for a transaction to store in a block), it can be identified too. |
| Block malicious vehicle | The RSU blocks the malicious vehicle after a threshold of transactions. Trust values are carried to other RSUs, and if a vehicle is blocked by any RSU, it cannot enter a new block of other RSUs. |

A malicious vehicle creates an impersonation attack by broadcasting a false message of a traffic crash with the forged digital signature. The system tries to block malicious vehicle identification through the digital signature with the majority votes of the stakeholders. In this way, a blockchain-based VANET blocks a "legitimate" vehicle as a malicious vehicle. So, with quantum computers [20], if we can figure out and break the encryption, such an attack would be possible [38]. An attacker gains the trust of the system, gets miner responsibilities, and verifies malicious transactions as authentic messages. Thus, the consensus of majority voting fails to remove the malicious vehicle as it has more stakes in the system by adding and verifying malicious vehicles as legitimate vehicles (increasing trust values) and using their votes to influence the system. The attacker starts double spending the initial gas amount at the beginning and spams the VANET with more false messages.

The PoW mechanism also acts on the consensus mechanism using the consensus mechanism in an unauthenticated Mempool [35]. A malicious vehicle can report the same traffic crash with other vehicles' digital signatures and authenticate the message. An attacker could have 10 minutes to either figure out how to crack the puzzle to act as a miner using Grover's algorithm [29] or the digital signature using Shor's algorithm utilizing quantum computers of other vehicles to spread the traffic crash notification from the Mempool.

## IV. IMPERSONATION ATTACK SCHEME

Our attempt to breach the blockchain's trust mechanism originates from the vulnerabilities associated with asymmetric cryptography, or public-key cryptography method. In asymmetric cryptography, a sender employs their private key (SK) to sign a transaction intended for recording on the blockchain, generating a digital signature. Subsequently, the sender broadcasts the transaction along with their public key (PK). The public key serves as a means for receiving nodes to validate that the digital signature originates from the rightful person. Asymmetric cryptography relies on either prime number factorization or the discrete logarithm problem to generate these key pairs. If one can break this algorithm with Shor's algorithm, it is possible to forge the private key and hence forge the digital signature. With the forged private key or digital signature, a sender (i.e., a malicious vehicle) broadcasts false messages or transactions to other legitimate vehicles and corresponding RSU within its communication range in the blockchain. These private-public key pairs are formed using the prime factorization in RSA. By using Shor's algorithm implemented in a quantum computer, a malicious vehicle can guess the prime factor and figure out the private key, thus forming the attack scheme.

This section presents an attack scheme that will be used to launch an impersonation attack using Shor's algorithm implemented in a quantum computer on a blockchain-based VANET. In this scheme, the trust chain's dependency on the corresponding trust value is exploited to detect any discrepancy by verifying the digital signatures and messages. Figure 2 presents an impersonation attack scheme against trust-based architectures using quantum Shor's algorithm, which is a sub-process to forge the digital signature and use it to broadcast a false message (e.g., traffic crash) with a forged digital signature. An attacker checks if the system is PoW-based architecture or not, as it must decide if there is a time constraint (key refresh time and block interval) for the attack to be successful. An attacker with a forged digital signature sends out false messages and gains trust by pinning its blame on legitimate vehicles using their forged digital signatures. As a malicious vehicle takes on the miners'/validators' responsibilities, it will improve the trust values of new malicious vehicles and help to get a majority in voting. A double-spending attack happens when it sends multiple transactions with the same gas value to save up its transaction limits (see Table I for more details).

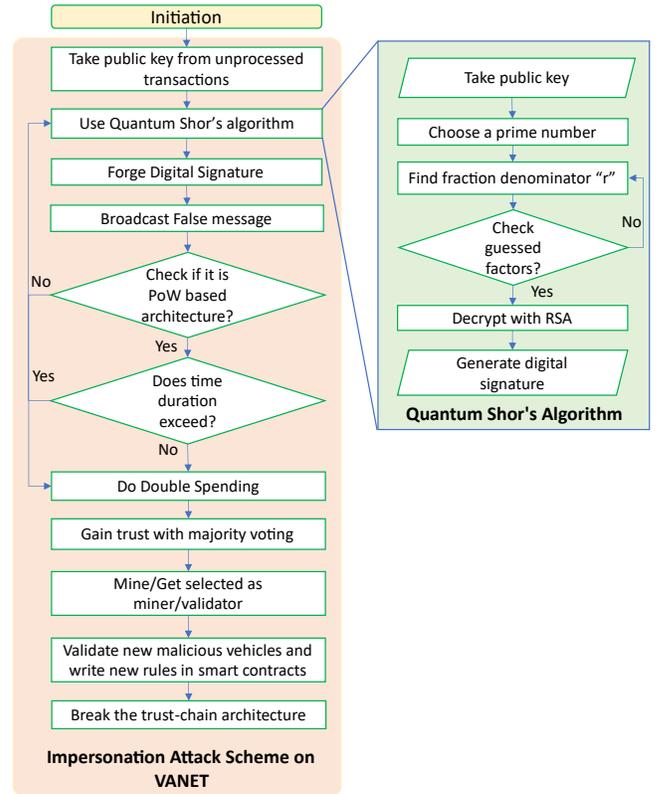

**Fig. 2.** Impersonation attack scheme using Shor's algorithm implemented in a quantum computer for blockchain-based VANET.

As a subprocess, Shor's algorithm implemented uses period finding (phase) in superposition states to perform modular exponentiation in polynomial time (see Figure 2) to break RSA. Note that Shor's algorithm uses the Quantum Fourier Transform (QFT) to find the period of a modular multiplication function efficiently. The modular multiplication function takes two integers, say $m$ and $x$, and computes the remainder of the product of $m$ and $x$ divided by some modulus, $N$. In other words, it computes $m^x \bmod N$. To factor a composite number $N$ using Shor's algorithm, we first choose a random prime number $m$ between $1$ and $N-1$. Then we apply the modular multiplication function repeatedly to compute $m^x \bmod N$ for a series of values of $x$ until we find two values of $x$, say $x_1$ and $x_2$, such that $(m^{x_1} \bmod N) = (m^{x_2} \bmod N)$. The period of the modular multiplication function is then given by the difference between $x_1$ and $x_2$, i.e., $p = x_1-x_2$. Finding the period, $p$, is the key step in Shor's algorithm, and it can be done efficiently using QFT. The QFT is a quantum version of the classical Fourier transform, which is a mathematical tool for analyzing periodic functions. By applying the QFT on the superposition states corresponding to the outputs of the modular multiplication function, the algorithm can find the period [23]. If the phase is odd or check $m^{\frac{r}{2}}$ ($m$ is a prime number) and it is congruent to 1 (mod $N$, where $N$ is a public key), it is in the order of $r/2$ ($r$ is a fraction denominator). Then, the $GCD(m^{(\frac{r}{2}-1)}, N)$ and $GCD(m^{(\frac{r}{2}+1)}, N)$ are applied to obtain prime factors of $N$.

## V. EXPERIMENTAL SETUP AND IMPLEMENTATION

In order to validate the effectiveness of the impersonation attack using Shor's algorithm implemented in a quantum computer, the performance evaluations of a blockchain-based VANET are conducted utilizing a simulation platform, i.e., Objective Modular Network Testbed in C++ (OMNET++) [36], [40].

### A. VANET Setup

For implementing this simulation platform, the simulation environment is divided into three modules: mobility, VANET, and quantum computing, as shown in Figure 3. To set up the mobility module, we used Simulation for Urban Mobility (SUMO) [41], which interacts with VEINS [42] and contains the road network and vehicle trajectory information. Overall, we simulated blockchain-based VANET using OMNET++, extended INET [43] library for V2V and V2I communication, as well as VEINS and SUMO for vehicular mobility. VANET has been tested and implemented using OMNET++ 5.6.2 version with the help of SUMO.

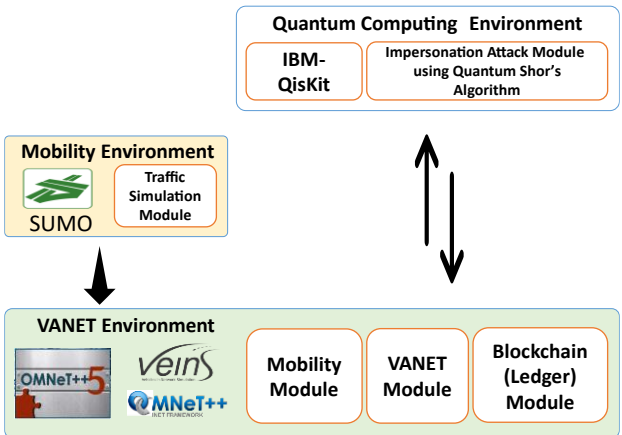

**Fig. 3.** Block diagram of system interaction architecture.

The VEINS (5.21 version)-SUMO integration has been done to set up the simulation of a traffic crash scenario. INET (version 4) is used to implement V2RSU (V2I) and V2V communication. Our experimental scenario is limited to four vehicles and a RSU to serve as proof of concept. In this setup, the attack propagates throughout the entire simulation due to the interoperability of the trust chain facilitated by message passing through a RSU. The vehicles provide consensus, while the RSU acts solely as a local administrator without participating in the voting process. Therefore, this constraint on network size does not impede the validity of the proof of concept. We implement the blockchain using C++ in OMNET++ with five nodes, where four moving nodes act as vehicles and one static node, i.e., RSU. In the omnetpp.ini file, we declare the simulation conditions, initial money, data rate, connectivity protocols, delays, number of malicious vehicles, and number of malicious transactions (acts as fault-tolerance limit, considered two in our simulation). In the network (.ned) file, we declare all the V2V and V2I communication protocols. In the vehicle files (vehicles and RSU), we define the message relay, packet generation (packet length, packet type, digital signatures, timestamp, hashed messages), receive, acknowledgment, block creation request, verification, and routing information. The success of the trust chain depends on successfully decrypting, validating, and verifying authenticated messages and finding malicious transactions and suspected vehicles at any point in time. We generate log messages (timestamp and trust value) for every transaction within the blockchain-based VANET. The RSU aggregates the voting of every vehicle to make malicious vehicle identification decisions as per the consensus mechanism.

As illustrated in Figure 4, after downloading the blockchain module and joining the network, every vehicle's onboard unit (OBU) has three modules: mobility, VANET, and local blockchain ledger. The mobility module consists of the roadway network and the vehicle's trajectory information. The VANET module enables V2V and V2I communication. The blockchain ledger module consists of records of transactions that contain a digital signature, a message, a timestamp, and transaction value details. Each vehicle uses the blockchain ledger module to send votes for each transaction. The RSU has the VANET module as it receives messages and disseminates transaction messages. It also contains a ledger module, which checks authenticity using majority voting, and maintains a list of suspected vehicles and their corresponding block lists.

We use VEINS and SUMO to generate vehicle trajectory information. A dedicated ledger module is coded in OMNET++, and INET is used for VANET communication. For each message, a sender (a vehicle or an RSU) generates a packet, encrypts it with RSA using an asymmetric key, and broadcasts it to a receiver (a vehicle or a RSU). This communication between a sender and a receiver will be stored in their respective ledgers. This ledger information consists of their transaction value, which is a gas value as defined above, their digital signature, message, and timestamp. If both sender and receiver are vehicles, these transaction details are also sent to the RSU by the receiver, and the RSU disseminates this transaction to other vehicles to maintain consistency in the ledgers of other vehicles. Each vehicle generates the creditability of this transaction and uploads its trust value using either "1" or "0." Here, "1" represents a trusted transaction, and "0" represents a non-trusted transaction.

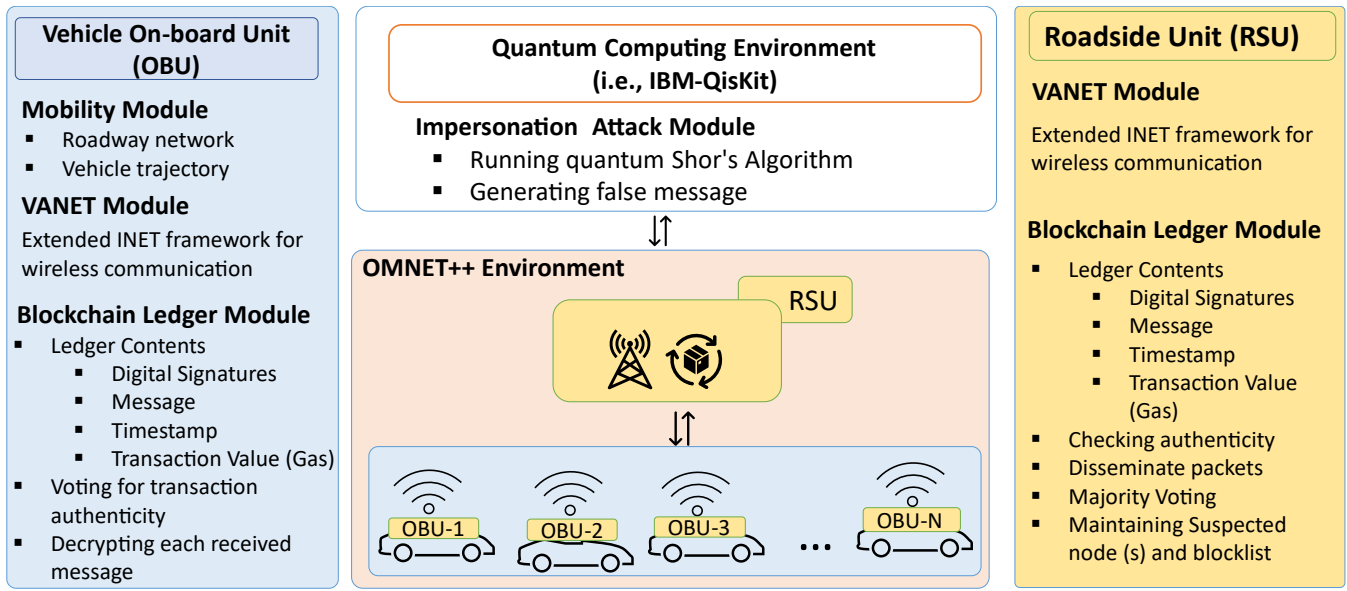

**Fig. 4.** Implementation of impersonation attack model utilizing Shor's algorithm implemented in a quantum computer.

The RSU then checks the transaction's digital signature and corresponding gas value for double spending scenarios. In addition, the RSU checks the overall authenticity of the transaction, and the corresponding message based on majority voting. The trustworthiness of each transaction is decided based on a trust threshold value. If trusted, it increases the trust value of the overall system; otherwise, it puts the vehicle on a suspected list. At least two transactions are needed to identify a malicious vehicle as fault tolerance, block it, and pass it to other RSUs. The transaction value (or gas spent in every transaction) and trust value are calculated. The key parameters related to Shor's algorithm-based attack model are provided in Table II.

TABLE II
KEY PARAMETERS

| Parameters | Values |
|---|---|
| Number of vehicles | 4 |
| Number of RSU | 1 |
| Packet size | 100 bytes |
| Encryption type | RSA |
| Hash algorithm | SHA-256 |

*B. Blockchain Setup*

Incorporating the trust chain architecture into OMNET++, the system features a non-blocking transaction system designed for efficient communication among agents, supporting multiple simultaneous transactions. Each agent owns and maintains their data, establishing local trust levels for scalability and stable performance. Every transaction creates a new block encompassing transaction data, previous block information, sequence numbers, public keys, and digital signatures. Initiated by the genesis block, the ledger progresses with subsequent blocks and agreed-upon transactions (using the consensus mechanism), ensuring transparency and security in a decentralized network.

In a network of five nodes, including one static RSU and four vehicle nodes, trust-chain facilitates transactions through four operations in each vehicle: (i) transaction request (TR), (ii) chain request (CR), (iii) chain replay message (CRM), and (iv) transaction confirmation (TC); and four operations in each RSU: (i) transaction verification (TV), (ii) dissemination (DM), (iii) chain verification (CV) and (iv) chain confirmation (CC). The initialization step sets up parameters, initializes structures, and registers signals for logging. Incoming messages initiated by each vehicle go through the transaction request (TR) process. Then, the messages are processed by RSU with the transaction handling, taking various actions such as transaction verification (TV), dissemination (DM), chain verification (CV) and chain confirmation (CC) based on message types like transaction or chain requests.

To ensure blockchain integrity, local and global logs are maintained in the log database, preventing duplicate entries. The RSU validates received transaction information against the global chain, and the vehicle checks its validity or duplicity against the local chain. The RSU oversees the successful dissemination of messages to vehicles through acknowledgement. At the TV step, the RSU checks authentication of messages. The system fortifies against malicious nodes through various functionalities, including randomized hashed addresses in the CR process, verification of digital signature in TV, and setting up public-private keys in initiation. With the voting phase in TC, the local and global trust levels are calculated, determining the threat levels of nodes. To ensure a message doesn't sit on the unprocessed transactions indefinitely, the chain records the start and end timings of transactions and discards the request after 10 minutes. Note that the simulation stops if malicious transactions by an entity (vehicle or RSU) exceed a threshold number of malicious transactions.

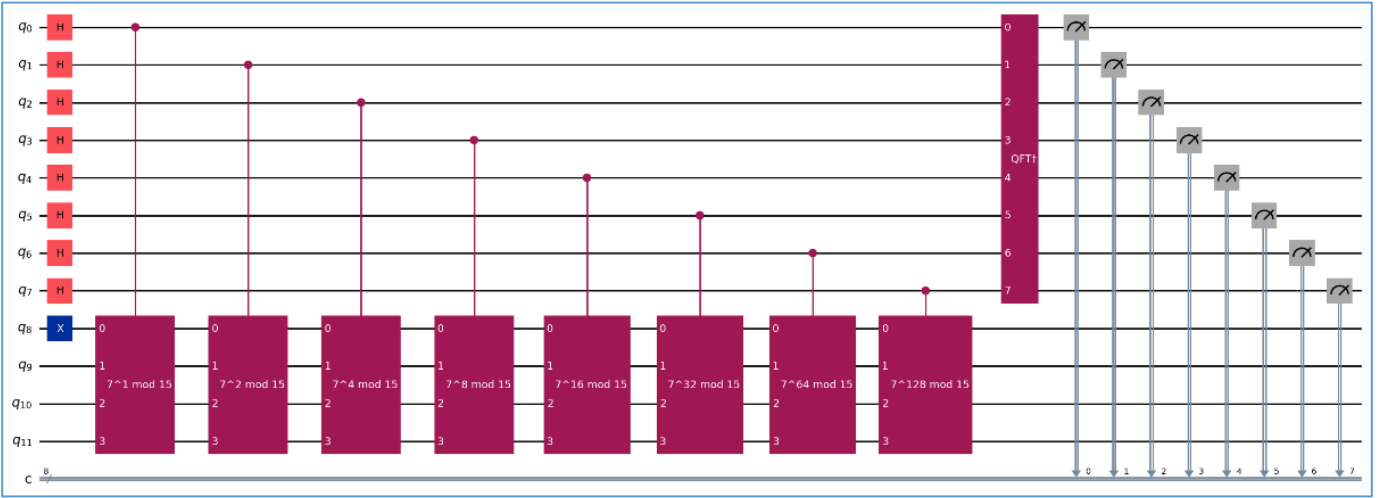

**Fig. 5.** Quantum circuit implementation of modular exponentiation for this study using IBM Qiskit.

Trust-chain's gas values, equivalent to the concept of transaction fees in Ethereum (ETH), are dynamically calculated based on the complexity and resources required for transactions. Unlike Ethereum's fixed gas fees, trust-chain employs a flexible gas pricing model that adjusts based on network demand and computational intensity. This approach ensures cost-efficient transactions while accommodating varying computational loads. In this simulation, we considered a gas value of 100 ETH. Since an Ethereum transaction typically consumes 21,000 units of gas and the base fee is established at 10 gwei (a smaller unit of Ether), the calculation for gas per transaction amounts to (21,000 * 10) = 210,000 gwei [transaction fee = (gas unit * (base fee per gas + max priority fee per gas))] as per [44], which is equivalent to 0.00021 ETH. The maximum priority fee is assumed to be 0 for this scenario. As a result, this simulation has the capacity to accommodate a number of transactions within the specified fee parameters to accommodate this simulation. The adversary detection model assumes any node may be malicious, with trust-chain tested under adversarial conditions involving false message injection and double-spending attacks at intervals to evaluate performance.

### C. Quantum Computing Environment Setup

In our quantum simulation, we employ a small key RSA-based message encryption implemented in the Qiskit software development kit (SDK), which is part of the IBM Quantum environment [35]. RSA-based encryption utilizes prime factorization through Shor's algorithm to forge digital signatures of neighboring vehicles. The quantum Shor's algorithm is applied to decipher the digital signature of the sender, allowing the generation of false messages with forged digital signatures for creating an impersonation attack within the VANET. The IBM Quantum Lab [24] is utilized for conducting this simulation experiment.

For our experiment, we used the quantum simulator (Aer Simulator) to execute the circuit on IBM's actual quantum architecture. The factorization relies on the Quantum Phase Estimation (QPE) algorithm, which iteratively determines the phase, which is then converted into a fraction to deduce the period $r$. Based on $r$, it calculates potential factors of $N$ using guesses and validates them. Initially, Hadamard gates are applied to initialize the qubits into a state of superposition. Subsequently, controlled unitary operations are used to execute modular exponentiation

The QFT is then applied to the qubits to extract the phase. Upon extracting the phase, the inverse QFT operation is performed using a combination of Hadamard gates and controlled phase rotation gates. Once these operations are completed, the quantum circuit is compiled into a basis gate set suitable for simulation on quantum hardware. The circuit is then executed using a quantum simulator, specifically the AerSimulator.

Utilizing the Qiskit SDK, we can efficiently determine the co-prime of a small prime number (15) capable of encrypting any number between 0 and 9 as a digital signature. Moreover, it can encrypt up to any of 14 sequential letters out of the 26 alphabets in English, serving as messages in V2V communications. This capability allows us to break the small key RSA algorithm using Shor's algorithm. Subsequently, we generate messages with forged digital signatures from a vehicle chosen as the malicious entity in our simulation. The quantum attack compromises the entire trust mechanism of the blockchain, as it incorrectly identifies another vehicle as malicious while failing to pinpoint the original malicious vehicle. Consequently, the trust value of the falsely accused vehicle increases, leading it to be selected as the miner in the blockchain network. The malicious vehicle, now acting as the miner, introduces more fraudulent transactions from other vehicles into the new block, further jeopardizing the integrity of the VANET. This simulation demonstrates the vulnerability of the trust mechanism against an attack model implemented in a quantum computer, emphasizing the importance of quantum-resistant cryptographic methods in securing V2V communications in a blockchain-based VANET.

Figure 5 depicts the quantum circuit implementation with 8 qubits($q$) and 4 classical bits ($c$) for the prime factorization of 15, where $m=7$ and $N=15$. Four of the counting qubits ($q_0, q_1, q_2, q_4, q_5, q_6$ and $q_7$) are into superposition states of ($|0>, |1>$) with equal probability using the Hadamard gate (see "*H*" in Figure 5). Then, we use *X*-gate or Pauli-*X* gate (see "*X*" in Figure 5) to convert qubit($q_8$) from State-0 (an auxiliary qubit) to State-1. After that, *m* is raised to the power of $2^q$ modulo *N* using a controlled operation to find the modular exponentiation. Later, all gates are appended to aggregate every qubit, *q*, in each iteration of modulo operation to form the circuit. After that, an inverse QFT is performed on the counting qubits to determine the period finding value (state $|s>$) from the frequency domain (superposition state $|t>$) and appended (with every qubit *q*) to get the circuit. Then, the counting qubits are measured and stored the results in the classical bits.

## VI. EVALUATION OUTCOMES

This section is structured into five subsections. The first subsection focuses on setting up the simulation platform by showing blockchain based VANET communication characteristics with the simulation time. The second subsection discusses the evaluation of the failure of a trust-based PoS-VANET system if it has been affected by an impersonation attack that uses a quantum computer. The accumulated trust value was calculated by adding offsets in every triggered event and by generating event logs. In the third subsection, we evaluate the failure of the trust mechanism due to an impersonation attack by examining each vehicle profile. The fourth subsection assesses the computational time required for a quantum computing-powered attack to meet the time-limited consensus-based VANET's blockchain architecture. Finally, we analyze IBM's existing roadmap concerning the required number of noisy qubits for the factorization of RSA-2048 bits.

### A. Blockchain Data Generation and Communication Characteristics without Impersonation Attack

In this subsection, we first generated blockchain data logs to assess the communication characteristics that established the framework for a thorough analysis of attack scenarios. This subsection helps to evaluate the system performance and throughput of the architecture in a non-attack scenario. Generating such simulation data by logging and calculating trust values ensures system security and integrity. During the phase of system setup, a public-private key pair is generated, and hash functions required for digital signatures are created to maintain security. Data logging starts with tracking and analyzing initial transactions, the dissemination and reception of messages, and the state of trust chain. These data logs contain detailed repositories of transactional information in the simulation, including timestamped events detailing the routing of packets and information about source, destination, and transaction amount. The logging process also shows transaction acknowledgement of packet transmission and reception, providing information about the start and end of data transfer between vehicles and RSUs. The RSU disseminates messages about the transaction. These dissemination messages provide packet propagation across the network and hold details, such as timestamps, transaction amounts, and sender and recipient signatures.

Logging also includes trust value calculation details, such as voting messages for consensus from vehicles. Finally, trust chain logs work as comprehensive repositories for trust value statistics and transaction information, essential for ensuring system security and validating transaction integrity. For instance, at a particular timestamp, both global and local trust-chain provide insights into trust chain behavior and initiates validator/miner chain requests and chain logs during packet handling and transmission.

From the logging of messages, we observe the vehicular and RSU communication over the simulation time, such as transmitted messages, chain request messages, disseminated messages, voting messages, and transaction acknowledgement (Ack) messages. Vehicle authentication and message verification with voting results and dissemination messages increase communication overhead with simulation time. Figure 6 depicts a segment of the logging of the simulation being conducted over a time span from 0 to 7.003 seconds, with the transmission starting at 5.180 seconds. During this period, messages were transmitted and received from the RSU and four vehicles. Within the simulation time, RSU transmitted 32 messages, which includes disseminated messages from RSU to all vehicles: Vehicle#1 transmitted 8 messages, Vehicle#2 transmitted 11 messages, Vehicle#3 transmitted 10 messages, and Vehicle#4 transmitted 10 messages. Analyzing the reception of messages, the RSU received 1 message within this time period (as there was no voting done, no voting messages were exchanged, and only one chain request was made to RSU), Vehicle#1 received 9 messages, Vehicle#2 received 12 messages, Vehicle#3 received 9 messages, and Vehicle#4 received 10 messages. This results in an average transmission rate of approximately 10.42 messages per 100 seconds and an average reception rate of approximately 8.5 messages per 100 seconds. RSU has a higher transmission rate as it is responsible for disseminating messages.

It is worthy to note that the throughput of the communication network—both in non-attack and attack scenarios—remains relatively unchanged due to the nature of the impersonation attack, which does not directly alter the volume of messages, or the data transmission rates within a network. This characteristic makes the impersonation attack particularly difficult to maintain the integrity of messages through traditional blockchain-based system monitoring and throughput analysis alone. The attack seamlessly integrates within the trust-based framework, maintaining typical throughput levels while subtly compromising trust integrity through forged digital signatures. This highlights the importance of enhanced security mechanisms beyond mere throughput analysis to detect such intelligent attacks.

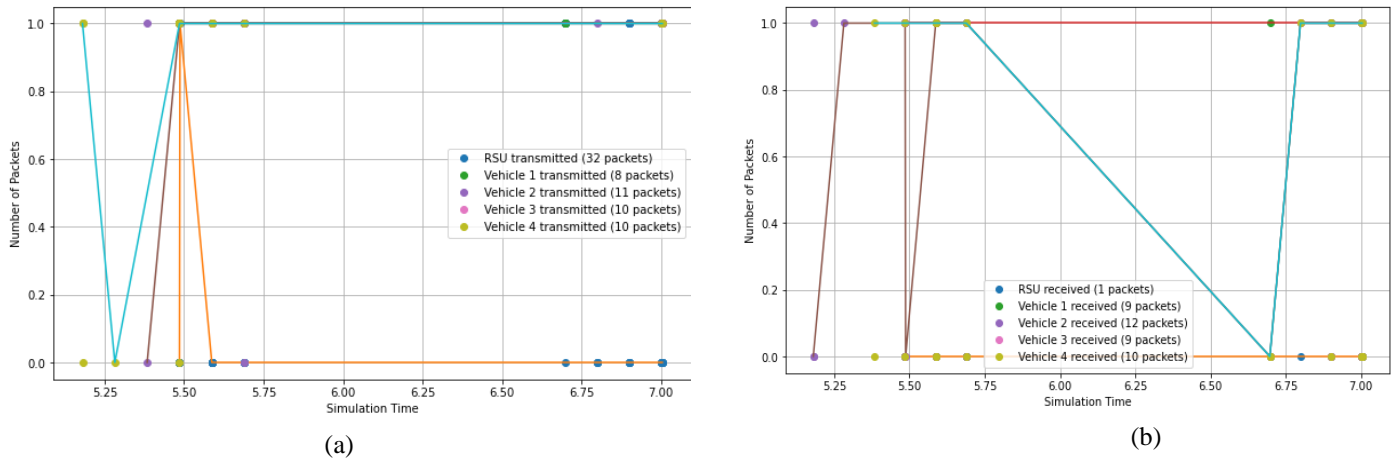

Fig. 6. No of packets vs simulation time over a time span from 0 to 7.003 seconds (a) data transmission, (b) data reception.

### B. Evaluation of Trust-based System Failure

We present a trust-based VANET architecture using simulation parameters (as shown in Table I) that fails to detect and increases the trust value of a malicious vehicle (i.e., Vehicle #2). Note that Vehicle #2 impersonates a legitimate vehicle, i.e., Vehicle #3, to complete a malicious transaction. As shown in Figure 7, we observe that the accumulated trust value in the RSU increases even if a malicious transaction or impersonation attack creates false messages that occur within the VANET. Trust value increases if the blockchain finds a malicious vehicle responsible for the false message transaction. However, this creates controversy on the trust-based VANET system, as it cannot decrease the trust until it finds a malicious transaction that has been impersonated and compromised by a malicious vehicle itself. This proves the vulnerability of the trust-based system. We also observe that a new block has been written by the validator, even if it is a malicious transaction, which leads to more malicious transactions in the VANET.

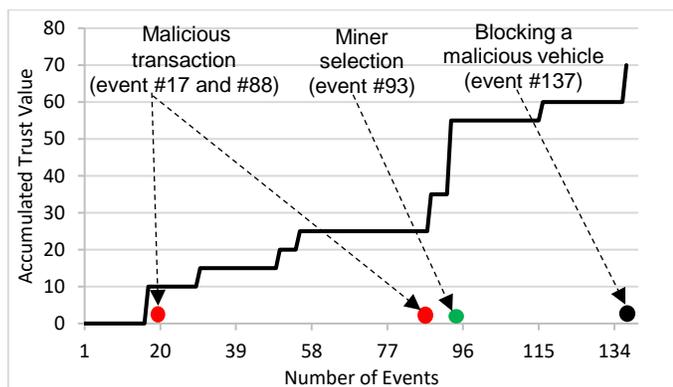

Fig. 7. Variation of accumulated trust value in RSU.

We found (see Figure 7) that the accumulated trust value increases for an event with an increment of 10 even if a malicious transaction occurs (indicated as red circles). On the other hand, the accumulated trust value increases for an event with an increment of 20 due to mining a new block (see the green circle), whereas the trust value increases with an increment of 5 for an authenticated transaction. This shows how a trust-based architecture fails to resist an impersonation attack and assigns mining responsibilities even if a malicious transaction occurs.

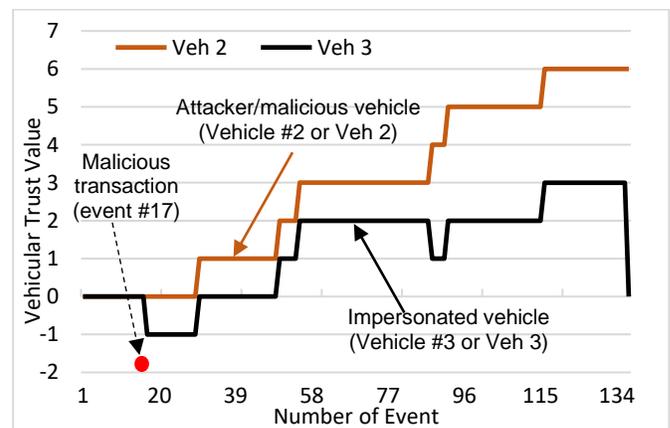

Fig. 8. Vehicular trust value profiles.

### C. Evaluation of Vehicular Profile of Trust Value

In this subsection, we evaluate trust-chain value of each vehicle at the vehicular level without considering trust offsets. Figure 8 illustrates the trust value profiles for Vehicle #2 and Vehicle #3. As Vehicle #2 executes as an impersonation attack utilizing a quantum computer on Vehicle #3, it forges the digital signature of Vehicle #3, causing the trust value of Vehicle #3 to decline for each malicious transaction. However, the trust value of Vehicle #2 increases, making it appear like a trusted node in VANET. We observe that Vehicle #3 gets blocked, and its trust value reaches zero when it appears to perform malicious transactions more than the threshold. Figure 8 shows that Vehicle #3 gets suspected of a malicious transaction by blockchain consensus mechanism at event 17, and the trust value reduces to -1. Vehicle #2 does this impersonation attack, but it does not get suspected due to its forging of digital signature, and its trust value increases to +1. Again, the same happens at events #88 and #137, where Vehicle #2 pins the blame on Vehicle #3, and the trust value of Vehicle #3 declines and eventually gets to 0 as it gets blocked. Vehicle #2, despite being the malicious node, becomes the miner at event #93 by highest trust values gets selected as the miner. This depicts the evaluation of individual vehicular level trust mechanism and failure of PoS-based trust-chain architecture (as stakeholders' consensus fails to identify the malicious vehicle).

### D. Computational Time Requirement

We assess the time required for generating a quantum-based private key to satisfy the computational time requirement for RSA [15]. To model an impersonation attack that uses a quantum computer, it is first necessary to break RSA, decrypt the received message, and forge digital signatures. Note that the private keys change with a predefined frequency of 10 mins as per PoW mechanism-based blockchain [6]. Our approach only takes polynomial time complexity and can identify the factors for a small key efficiently, which acts as a proof of concept. In our case, factoring a 4-bit number like 15 takes between 5 to 17 seconds. The experiment was conducted 30 times, as this sample size of 30 is considered significant for any central limit probability distribution. We developed this impersonation attack with a small key RSA (4 bits in our experiment) as proof of concept due to existing limited computational capacity to break 2048-bit RSA. Figure 9 shows that the mean factorization time is 10.4 seconds using a Poisson probability distribution of 30 trials. In addition to this factorization time, additional computational time is required to decrypt a message and create a false message packet with forged digital signature. The OMNET++ stores the total computational time (factorization time, message decryption and forged message creation) in a log file as shown below.

*Total Computational Time = Factorization Time + Message Decryption + Forged Message Creation*

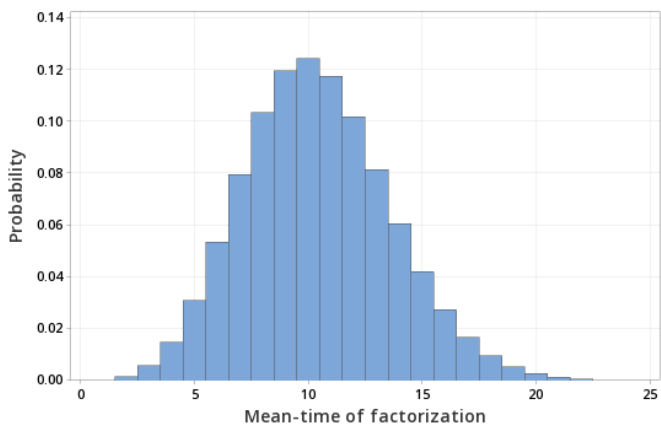

**Fig. 9.** Probability distribution of required time (in seconds) for factorization.

Overall, it takes a mean of 115 seconds, which is significantly lower than the 10 minutes required to hack into the current block. This ensures that the false message in "Mempool" gets sufficient votes within the time constraint.

Thus, our cyber-attack generation technique utilizes Shor's algorithm implemented in a quantum computer and takes advantage of the current limitation of blockchain-based VANET.

### E. Advancing Toward Computational Requirements to Overcome RSA-2048 Bits

We developed this impersonation attack with a 4 bits RSA as a proof of concept due to the existing limited computational capacity to break a 2048-bit RSA. For evaluating the feasibility of the impersonation attack utilizing Shor's algorithm implemented in a quantum computer, particularly for breaking cryptographic schemes like RSA-2048, it is necessary to show a roadmap toward achieving quantum speedup in the Noisy Intermediate Scale Quantum (NISQ) era. Shor's algorithm, a key component of our impersonation attack scheme, demonstrated its potential in 2019 by successfully factoring the number 35 [45]. For the application of Shor's algorithm in factorizing an n-bit integer, the quantum circuit, as outlined in [45], utilizes a total of $2n + 3$ qubits. These qubits are specifically allocated for various tasks, including modular addition, overflow prevention, and ancillary registers for modular multiplication. Applying this formula to 2,048-bit integer results in a requirement of 4099 qubits ($2*2048+3$). By analyzing IBM's quantum computing roadmap, which incorporates the dynamics of noisy qubits, our projection in Figure 10 suggests that between 2032 and 2044, IBM's quantum infrastructure may have the capacity to efficiently factorize RSA-2048 efficiently. This prediction relies on exponential and linear regression modeling.

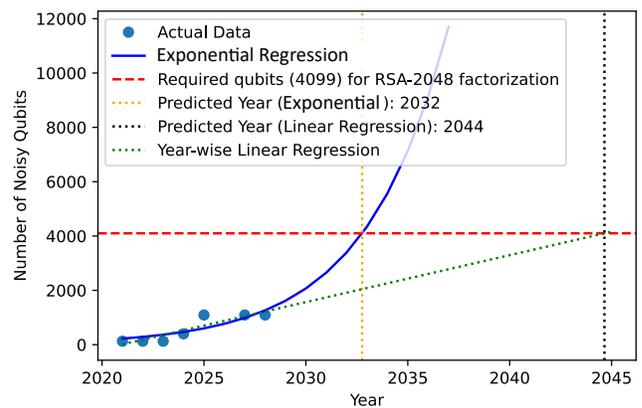

**Fig.10.** IBM quantum roadmap for factorizing RSA-2048 bits.

Although linear regression assumes a linear time-qubit relationship, which does not fully capture quantum computing's growth, it serves as a conservative baseline for comparison, representing a worst-case scenario. Despite efforts to predict quantum computing progress, there's a disparity between the number of qubits and their quality (fidelity), posing obstacles to predict the roadmap towards factorization of 2048-bits. While companies like D-Wave and IBM have experienced exponential growth in qubit numbers which is greater than its competitors, advancements in high-quality qubits characterized by extended coherence times and low error rates have progressed more slowly [46]. IBM utilizes "quantum volume" as a metric to assess the performance and quality of its quantum computers. Quantum Volume measures both the number of qubits and the depth of quantum circuits, offering insights into the system's capabilities. The IBM quantum roadmap for factorizing RSA-2048 bits shows the necessity of understanding quantum computing's current constraints, advancements, and potential future progress in securing cryptographic applications (see Figure 10).

## VII. LIMITATIONS OF QUANTUM BLOCKCHAIN AND POST-QUANTUM BLOCKCHAIN

Looking ahead, exploring quantum blockchain and post-quantum blockchain can make VANET more secure. By integrating quantum-resistant cryptographic algorithms into blockchain structures, we can better protect against potential threats using quantum computer, which is a practical and forward-thinking approach to VANET security. There could be two primary options to counter quantum threats to blockchains: quantum blockchains and post-quantum blockchains.

Quantum blockchains use quantum computing concepts like quantum key distribution (QKD), quantum digital signatures (QDS) and quantum teleportation to enhance efficiency and functionality. The QKD-based technique requires dedicated quantum communication channels [8], and physically setting up such channels is not feasible in a dynamic network, such as VANET. In addition, such long-distance QKD transfer in quantum secure direct communication channels, which is known as quantum relay, could have increased quantum error and complexity. Moreover, quantum relays [47] are prone to noise and errors as well as concurrent eavesdropping (compromised RSU in our case), which will be almost impossible to make a fault-tolerant and randomized (random qubit measurement) QKD [5] to address security issues. Quantum blockchains, while leveraging innovative concepts, face obstacles such as the feasibility of establishing dedicated quantum communication channels and mitigating inherent quantum error and complexity.

A few studies have been conducted on improving the ownership mechanism of blockchain and making it quantum-safe through post-quantum cryptography [48] and quantum key distribution (QKD) [13]. However, post-quantum solutions lack standardization, suffer from periodicity and symmetry, and use large-size keys, which increase the complexity of the decryption of the key, such as a lattice-based architecture [7], [49]. Hash-based cryptography [50] and multivariate cryptography [51] exhibit a drawback in the form of large signature sizes, leading to a larger block-size and consequently, larger memory size. Similarly, code-based cryptography encounters the issue of increasing complexity due to larger key sizes, demanding extensive memory storage, and the risk of decoding failures when utilizing smaller keys in specific scenarios [52]. While isogeny-based cryptography is relatively new and holds prospects, it requires further enhancements to be suitable for practical real-world use. This involves tackling issues related to efficient implementation and conducting thorough security analyses of specialized schemes such as SIDH (Supersingular Isogeny Diffie-Hellman) and CSIDH (Commutative Supersingular Isogeny Diffie-Hellman) [53]. Post-quantum blockchains can focus on signature approaches, privacy-enhancing techniques, and developing decentralized applications (dApps) using smart contracts to provide more robust chain security. Thus, both quantum and post-quantum blockchain solutions, despite their potential, are still in their early stages of development in effectively countering quantum-based attacks.

## VIII. CONCLUSION AND RECOMMENDATION

The impersonation attack model presented in this paper with Shor's algorithm implemented in a quantum computer demonstrates a proof of concept to expose the vulnerabilities of a blockchain-based VANET. Even though existing quantum computers do not have the necessary computational efficiency to break a 2048-bit RSA, we conclude that the presented impersonation attack on a 4-bit RSA could act as a proof-of-concept for such an attack. The paper showcases the feasibility of an impersonation attack, with a scale-down of the current blockchain encryption standards, for compromising the trust chain of the blockchain-based VANET, demonstrating the persistence of potential attacks if scaled up. The vulnerability in the trust-based framework of blockchain, particularly in VANETs, is formulated and examined by the potential for impersonation attacks, wherein malicious actors can forge digital signatures, undermining the integrity of the trust mechanism. The paper shows that quantum computing-based attacks could exploit these vulnerabilities, thus posing significant threats to VANETs reliant on blockchain technology. In this study, an impersonation attack model utilizing Shor's algorithm implemented in a quantum computer is developed against a PoS architecture along with the consideration of a time-constraint-based PoW mechanism, and its effectiveness has been evaluated. Specifically, the vulnerability of vertical trust, i.e., vehicle-to-infrastructure trust-chain, using the aggregated trust profile at RSU and the horizontal trust, i.e., vehicle-to-vehicle trust-chain, is evaluated using each vehicle's trust profile. It reveals that a quantum computing-based cyber-attack, i.e., an impersonation attack, is able to successfully break the trust chain of a blockchain-based VANET. We have also analyzed a timeline and discussed a roadmap outlining the progression of quantum computers required for such an attack to succeed. The blockchain-based VANET requires feasible post-quantum cryptography solutions that do not have periodicity, symmetry, or large key problems. In future studies, quantum-based short key based post-quantum solutions and quantum solutions such as quantum digital signatures (QDS), quantum key distribution (QKD) over quantum satellites, or 1000 km long quantum communication channels could be explored. Therefore, a quantum-secured, blockchain-based VANET could serve as a solution for ensuring secure vehicular networks in the future.

ACKNOWLEDGMENT

This material is based on a study supported by the USDOT National Center for Transportation Cybersecurity and Resiliency (TraCR) under grant number 2244371. Any opinions, findings, and conclusions or recommendations expressed in this material are those of the author(s) and do not necessarily reflect the views of the TraCR, and the U.S. Government assumes no liability for the contents or use thereof.

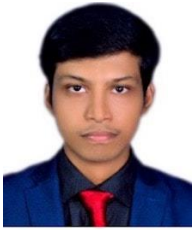

**Kazi Hassan Shakib** (Student Member, IEEE) received his B.Sc. degree in Computer Science and Engineering from Chittagong University of Engineering & technology (CUET). He received his M.S. in Civil Engineering from the University of Alabama, Tuscaloosa in 2023. Currently, he is a Ph.D. student at Department of Computer Science in Kansas State University. His research interests include Quantum Computing, Post-Quantum Cryptography, Internet of Things, and Encryption.

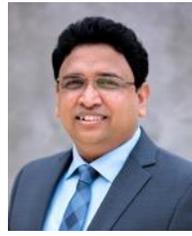

**Mizanur Rahman** (Senior Member, IEEE) received the B.Sc. degree in civil engineering from Bangladesh University of Engineering and Technology, Dhaka, Bangladesh, in 2008, and the M.Sc. and Ph.D. degrees in civil engineering (transportation systems) from Clemson University, Clemson, SC, USA, in 2013 and 2018, respectively. He is an Assistant Professor with the Department of Civil, Construction and Environmental Engineering, The University of Alabama, Tuscaloosa, AL, USA. His research focuses on cybersecurity in transportation cyber-physical systems.

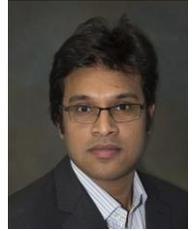

**Mhafuzul Islam** is currently working as a Senior Researcher at General Motors. He received his Ph.D. in Civil Engineering in 2021 from Clemson University. He also received a B.S. degree in Computer Science and Engineering from Bangladesh University of Engineering and Technology (BUET) in 2013 and an M.S. degree in Civil Engineering from Clemson University in 2018. His research interest includes transportation cyber-physical systems with an emphasis on data-driven connected autonomous vehicles utilizing machine learning.

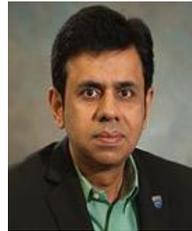

**Mashrur Chowdhury** (Senior Member, IEEE) received his Ph.D. degree in civil engineering from the University of Virginia in 1995. He is the Eugene Douglas Mays Chair of Transportation in the Glenn Department of Civil Engineering, Clemson University, SC, USA. He is the Founding Director of the USDOT sponsored USDOT National Center for Transportation Cybersecurity and Resiliency (TraCR). He is also the Director of the Complex Systems, Data Analytics and Visualization Institute (CSAVI) at Clemson University. Dr. Chowdhury is a Registered Professional Engineer in Ohio, USA. He is a Fellow of the American Society of Civil Engineers and a Senior Member of IEEE.